\newcommand{\msun}{\mbox{$M_{\odot}$}}

\newcommand{\lsun}{\mbox{$L_{\odot}$}}

\newcommand{\rstar}{\mbox{$R_{\ast}$}}

\newcommand{\teff}{\mbox{$T_{\rm eff}$}}
\newcommand{\Teff}{\mbox{$T_{\rm eff}$}}

\newcommand{\ratio}{\mbox{$v_{\infty}$/$v_{\rm esc}$}}
\newcommand{\mdot}{\mbox{$\dot{M}$}}

\newcommand{\msunyr}{\mbox{$M_{\odot}\,{\rm yr}^{-1}$}}
\newcommand{\rin}{\mbox{$R_{\rm in}$}}
\newcommand{\tin}{\mbox{$\mathrm{T}_{\rm in}$}}

\newcommand{\logll}{\mbox{$\log (L/L_{\odot})$}}

\documentclass[10pt,preprint]{aastex}
\usepackage{epsfig}

\begin{document}

\title{THE MISSING LUMINOUS BLUE VARIABLES AND THE BISTABILITY JUMP}

\author{Nathan Smith\altaffilmark{1}}

\affil{Center for Astrophysics and Space Astronomy, University of
Colorado, 389 UCB, Boulder, CO 80309; nathans@casa.colorado.edu}

\author{Jorick S.\ Vink}

\affil{Imperial College, Blackett Laboratory, Prince Consort Road,
London, SW7 2BZ, UK; j.vink@ic.ac.uk}

\author{Alex de Koter}

\affil{Astronomical Institute ``Anton Pannekoek'', University of
Amsterdam, Kruislaan 403, 1098 SJ Amsterdam, The Netherlands;
dekoter@science.uva.nl}

\altaffiltext{1}{Hubble Fellow}

\begin{abstract}

We discuss an interesting feature of the distribution of luminous blue
variables (LBVs) on the H-R diagram, and we propose a connection with
the bistability jump seen in the winds of early-type supergiants.
There appears to be a deficiency of quiescent LBVs on the S~Doradus
instability strip at luminosities between \logll$\simeq$ 5.6 and 5.8.
The upper boundary, interestingly, is also where the
temperature-dependent S~Doradus instability strip intersects the
bistability jump at about $\Teff \simeq 21\,000$ K.  Due to increased
opacity, winds of early-type supergiants are slower and denser on the
cool side of the bistability jump, and we postulate that this may
trigger optically-thick winds that inhibit quiescent LBVs from
residing there.  We conduct numerical simulations of radiation-driven
winds for a range of temperatures, masses, and velocity laws at
\logll=5.7 to see what effect the bistability jump should have. We
find that for relatively low stellar masses the order of magnitude
increase in the wind density at the bistability jump leads to the
formation of a modest to strong pseudo photosphere that might alter a
star's apparent position on the HR diagram.  The effect is strongest
for LBVs approaching 10 M$_{\odot}$, where the pseudo-photospheres are
sufficiently extended to make an early B-type star appear as a yellow
hypergiant.  Thus, the proposed mechanism will be most relevant for
LBVs that are post-red supergiants (curiously, the upper boundary at
\logll$\simeq$5.8 coincides with the upper luminosity limit for red
supergiants).  Further work is obviously needed, especially with
regard to a possible evolutionary connection between the ``missing''
LBVs and the most luminous red supergiants and yellow hypergiants.
Specifically, yellow hypergiants like IRC+10420 and $\rho$ Cas occupy
the same luminosity range as the ``missing'' LBVs, and show apparent
temperature variations at constant luminosity.  If these yellow
hypergiants do eventually become Wolf-Rayet stars, we speculate that
they may skip the normal LBV phase, at least as far as their {\it
apparent} positions on the HR diagram are concerned.

\end{abstract}


\section{INTRODUCTION}

The post-main sequence evolution of massive stars with $M_{\rm ZAMS}
\ga 30 M_{\odot}$ is very different from that of stars with lower
mass, and is still poorly understood.  An important difference
observationally is that massive stars evolve at nearly constant
bolometric luminosity, traveling back and forth across the upper
Hertzsprung-Russel diagram (HRD).  Unlike low-mass stars, two stars at
the same position on the upper HRD may be in different stages of their
lives, sometimes even on the same evolutionary track, so
observationally the situation can be misleading.  Many evolutionary
sequences have been proposed to explain a star's journey from the
main-sequence to the He-rich and mass-depleted Wolf-Rayet (WR) phase,
including transition phases when a star is classified as a blue
supergiant (BSG), red supergiant (RSG), yellow hypergiant (YHG), B[e]
supergiant, P Cygni-type star, Of star, WNL star, etc.\ (see reviews
by Chiosi \& Maeder 1986; Langer 1989; Langer et al.\ 1994; de Jager
1980).  Post-main-sequence evolutionary tracks vary depending on mass
and luminosity; for example, stars with \logll$>$5.8 may not become
RSGs (Humphreys \& Davidson 1994, HD hereafter; Chiosi \& Maeder 1986;
Maeder \& Meynet 1987; Stothers \& Chin 1994, 1999).  Perhaps the most
critical evolutionary phase -- when a star may shed a great deal of
mass in a short time period before becoming a WR star -- is when it is
classified as a luminous blue variable (LBV).

The term ``luminous blue variable'' (Conti 1984) refers to a specific
class of unstable stars, even though most of the upper left part of
the HRD would seem to qualify (see Wolf et al.\ 1999).  LBVs (also
called S~Dor stars or Hubble-Sandage variables) exhibit a particular
kind of instability, which causes a star to brighten at optical
wavelengths ($\Delta V \simeq$ 1--2 magnitudes) as a result of a shift
in the bolometric flux from the UV to the optical (see reviews by HD;
van Genderen 2001).  The trigger of these events is not understood,
but the furious mass loss and generally-unstable nature of LBVs
results because they have lost considerable mass already, while
evolving at constant L. This has increased their L/M ratio so that
they may be in dangerous proximity to an opacity-modified Eddington
limit in their outer layers (Appenzeller 1986, 1987, 1989; Lamers \&
Fitzpatrick 1988; HD; Ulmer \& Fitzpatrick 1998).

Here we focus on the behavior of ``normal'' LBVs in their quiescent
state between outbursts.  In quiescence, LBVs generally reside along
the S~Doradus instability strip, shown by the diagonal shaded area in
Figure 1 (see Wolf 1989; HD).  We point out a distinct lack of LBVs in
the range $5.6 \ga \logll \ga 5.8$, and -- more importantly -- we
postulate a link between these missing LBVs and the
``bistability jump'' in the line-driven winds of luminous stars (Vink,
de Koter, \& Lamers 1999; Lamers, Snow, \& Lindholm 1995).  In \S 2 we
review the basic ideas behind the bistability jump, and in \S 3 we
discuss how this jump may affect quiescent LBVs.  Then, in \S 4 we
investigate numerically whether our idea has a theoretical foundation,
and finally in \S 5 we speculate about consequences of our proposed
connection for general aspects of evolution atop the HRD.

\section{THE BISTABILITY JUMP}

Luminous early-type stars have optically-thin line-driven winds (Lucy
\& Solomon 1970; Castor, Abbott, \& Klein 1975; Cassinelli 1979;
Pauldrach et al.\ 1986; Kudritzki \& Puls 2000), and show a
discontinuity in their wind properties near spectral type B1, at
effective temperatures around 21\,000 K.  The ratio $v_{\infty}/v_{\rm
esc}$ drops by a factor of two, from $\sim$2.6 for stars on the hot
side, to $\sim$1.3 for stars with spectral types later than B1 (Lamers
et al.\ 1995).  This observed bifurcation is referred to as the
``bistability jump''.  The drop in velocity toward cooler temperatures
is accompanied by an increase in the mass-loss rate by a factor of
$\sim$5, and an increase in the wind's ``performance number'' $\eta =
\dot{M}v_{\infty}/(L/c)$ by a factor of 2 to 3 (Vink et al.\ 1999).
The jump in wind parameters is caused by an increase in the driving
effect of Fe~{\sc iii} lines below the sonic point as the effective
temperature decreases (Vink et al.\ 1999).

This mechanism may also apply to individual stars with spectral types
near B1.  Pauldrach \& Puls (1990) introduced the bistability
mechanism in an effort to describe the curious behavior and physical
state of P Cygni's wind.  P Cygni is an LBV that suffers minor shell
ejection episodes, perhaps due to excursions back and forth across the
bistability jump (Pauldrach \& Puls 1990; Lamers et al.\ 1985); if
this is true, P Cygni will be important for our discussion below.
Additionally, bistability may play a role as temperature varies with
latitude on a rapidly-rotating star with gravity darkening; Lamers \&
Pauldrach (1991) proposed the bistability mechanism as an explanation
for the creation of outflowing disks around B[e] stars (see also
Pelupessy et al.\ 2000).

Vink \& de Koter (2002) showed that the winds of LBVs are indeed line
driven, and that the bistability mechanism may play a role in their
unusual variability and consequent excursions across the HRD.  Since
LBVs are already unstable due to high L/M values, and since quiescent
LBVs have a range of temperatures from 10\,000 to 35\,000 K, we might
expect the bistability jump to impact the mass-loss behavior of LBVs
in their quiescent state as well, when they reside along the S Dor
instability strip.  In this paper, we propose that this is indeed the
case, as outlined below.

\section{THE MISSING LBVS ON THE S DOR INSTABILITY STRIP}

Figure 1 shows the positions on the HRD of well-studied LBVs in the
Milky Way and nearby galaxies, as well as some related stars; it
borrows from similar diagrams presented by Wolf (1989), HD, and de
Jager (1998), but includes additional stars and updated values from
the literature. LBVs generally reside in the shaded areas of the
diagram, either along the diagonal quiescent S Doradus instability
strip (Wolf 1989) or along the constant-temperature vertical stripe
while in outburst (see Davidson 1987).  These trends would not be
noticed if we included any and all stars in the upper HRD; instead we
have selected only relatively rare and unstable stars thought to be at
a particular evolutionary stage.  From the literature, we have limited
Figure 1 to include three types of stars: 1) confirmed LBVs that have
exhibited S Dor-type variability, 2) candidate LBVs known to have
spatially-resolved circumstellar
shells,\footnotemark\footnotetext{Many stars like Ofpe/WN9 stars are
sometimes considered LBV candidates based only on their spectral type,
but do not have visible circumstellar material.} and 3) cool
hypergiants that have exhibited dramatic changes in effective
temperature and spectral type over the last few decades.  We should
also note that a range of different values for temperature and
luminosity are sometimes given by various authors for the same LBVs at
quiescence; temperature values typically differ by a few thousand K,
and luminosities generally disagree by $\pm$0.1 dex, and sometimes
more.  We chose not to plot error bars in Figure 1 for clarity, and we
needed to make some judicious choices about which values to take from
the literature; in some cases suitable averages were adopted.
However, in general, these choices do not undermine the main points of
this paper.

Figure 1 seems to support the existence of the diagonal S~Dor
instability strip, where most LBVs reside when they are not at maximum
visual brightness.  It is often remarked (e.g., HD; de Jager 1998;
Stothers \& Chin 1996, 1999) that LBVs come in two flavors: the
``classical'' or high-luminosity LBVs (like AG Car, R127, an R143) and
the relatively low-luminosity LBVs (like HR Car and R71).

However, one might choose to characterize this bifurcation somewhat
differently -- here we point out a distinct {\it gap} in the otherwise
continuous S~Dor instability strip.  In particular, {\it no confirmed
LBVs are observed with bolometric luminosities between \logll=5.6 and
5.8}.  An obvious but intriguing coincidence is that the upper
boundary of this gap at \logll=5.8 is also the observed upper
luminosity limit for RSGs and YHGs on the right side of the HRD (e.g.,
HD).  One might surmise a possible connection between the luminosity
gap on the S~Dor instability strip and this upper luminosity limit;
namely, that stars below \logll=5.8 for some unknown reason can evolve
to the red side of the HRD, and hence do not become normal LBVs --- or
eventually become LBVs with higher L/M ratios than they would
otherwise have had.  It is not the purpose of this paper to
investigate why these less luminous stars can become RSGs (see Lamers
\& Fitzpatrick 1988; Ulmer \& Fitzpatrick 1998).  Instead we offer an
independent potential explanation for why stars in this luminosity
range might be inhibited from appearing on the S~Dor instability strip
--- a reason to expect the apparent gap below \logll=5.8.  Possible
implications are discussed later in \S 5.

Since the S Dor instability strip is diagonal, one might wonder if the
``gap'' below \logll=5.8 is really a luminosity effect, or if it is
instead linked more directly to the characteristic {\it temperature}
of the star or opacity in its wind.  In this context, one can see that
the upper boundary of this gap at \logll=5.8 {\it is precisely where
the S~Dor instability strip crosses the bistability jump}, observed to
occur in the winds of blue supergiants at temperatures around 21\,000
K.  Is this just a coincidence, or is it possible that the bistability
mechanism is important in accounting for the ``missing'' quiescent
LBVs?  To the upper left of this critical juncture, classical LBVs
have powerful line-driven stellar winds and are observed to suffer
occasional outbursts because they are already unstable, perhaps due to
their proximity to the opacity-modified Eddington limit.  As we move
down the instability strip, how should we expect LBVs to behave as we
approach and then cross the bistability jump where the mass-loss rate
is expected to increase and the winds are expected to suddenly become
slower and more opaque?  It is instructive to consider the observed
behavior of LBVs and related stars near this critical juncture at
\logll=5.8 and $T \approx 21\,000$ K.

$\bullet$ P Cygni, for instance, is just above this limit.  Its wind is
unstable to small perturbations, and it was after studying P Cyg's
wind that Pauldrach \& Puls (1990) first proposed the bistability
mechanism.

$\bullet$ S Dor and Var C are also just above the critical luminosity
of $\logll \simeq 5.8$, and their apparent temperatures in quiescence
place them very close to the bistability jump. S Dor is unusual
compared to other LBVs in that it is more often observed in the
cooler, maximum-light phase and is rarely seen in the hotter quiescent
phase (HD; Wolf 1989).

$\bullet$ W243 is a newly-identified LBV in the cluster Westerlund 1
(Clark \& Negueruela 2004).  A recent spectrum of this star is almost
identical to IRC+10420, implying that it may have formed a cool pseudo
photosphere, although this is not a unique interpretation.

$\bullet$ No LBVs are seen just below $\logll \simeq 5.8$, but several
unstable YHGs exist in the same luminosity range as the hypothetical
gap on the S~Dor instability strip, such as the famous stars IRC+10420
and $\rho$~Cas (Humphreys et al.\ 2002; de Jager 1998).  These stars
have high mass-loss rates and dense stellar winds with unstable
convective atmospheres. Qualitatively like LBVs, the YHGs exhibit
curious changes in apparent temperature at constant bolometric
luminosity on timescales of years to decades. The lower boundary to
the LBV gap is less well-defined, but there seem to be plenty of LBVs
below $\logll \simeq 5.55$, while there are no known YHGs below this
luminosity.  All the YHGs are on the cool side of the vertical stripe
(at $T \sim 8\,500$ K) that marks the position of LBVs at maximum
light.  Is there a connection between the YHGs and the ``missing''
LBVs (de Jager 1989)?  Is there a connection with the so-called
``Yellow Void'' proposed by de Jager? We offer some motivated
speculation in \S 5.

$\bullet$ All but one of the {\it candidate} LBVs with shells found
within the luminosity range of the LBV gap are found on the hot side
of the bistability jump, even though properties of the nebulae around
candidates are similar to those of confirmed LBVs (e.g., Clark et al.\
2003a; Smith 2002; Pasquali et al.\ 1999).

Let us make the cautionary remark that LBVs are rare objects, and
consequently, that Figure 1 suffers from low number statistics.  So,
{\it is the LBV ``gap'' at $5.6 \la \logll \la 5.8$ real?}
Uncertainties in L are typically $\pm$0.1 dex, which would be enough
to make the gap quite dubious --- but to do so, errors would need to
conspire in such a way as to systematically lower the luminosities of
the classical LBVs, and to systematically raise the luminosities of
LBVs below $\logll \simeq 5.6$.  For a few objects, uncertainties in L
and T are considerably worse because of assumptions like the adopted
distance or reddening.\footnotemark\footnotetext{For example, some
authors have given a luminosity for HR~Car as high as $\logll \approx
5.7$ or 5.8 (for a large distance of 5.4 to 6 kpc; Shore et al.\ 1996;
Hutsemekers \& van Drom 1991), placing it within the LBV gap that we
have described here, although lower values closer to $\logll \simeq
5.5$ are often given as well.  These higher luminosities are somewhat
problematic, though, because then HR~Car's relatively low temperature
at quiescence would not fit into the S Dor instability strip.}  On the
other hand, the deficit of LBVs also holds for extragalactic LBVs,
where the distances (and bolometric luminosities) are more reliable
than in our own Galaxy.  In any case, it is plausible that some LBVs
may eventually be found to reside within the ``gap'', and the present
uncertainties do not yet support a claim that the gap is a pure {\it
void} with a complete absence of any LBVs.  Instead, we tentatively
interpret the gap as signifying a real {\it deficiency} of LBVs ---
{\it a location in the HRD where LBVs are less likely to be seen}.
Conditions in their atmospheres/winds make this difficult, but perhaps
not impossible as it also depends on factors like their L/M ratio,
rotation, and metallicity.  Nevertheless, observations (Figure 1) do
seem to support a deficit of LBVs at $5.6 \la \logll \la 5.8$.  This
is especially true when we consider that for any normal initial mass
function, LBVs should be more numerous at lower L, since LBVs extend
to luminosities as low as $\logll \simeq 5.4$.

Thus, {\it we propose that the LBV gap just below $\logll \simeq 5.8$
occurs in part because that is where the S~Dor instability strip
crosses the bistability jump.}  The increased mass-loss rates and
lower wind velocities expected on the cool side of the jump may
prevent these already-unstable stars from existing there; they may
have persistently dense winds with cooler pseudo photospheres, for
example.  This hypothesis is investigated numerically in the following
section, where we constrain the conditions required for it to apply.
Although various comments we made above concerning the two flavors of
LBVs and the apparent gap or division at $\logll \simeq 5.8$ might
have been noticed before and discussed in different terms (e.g., HD,
de Jager 1998; Humphreys et al.\ 2002), we believe that the specific
connection with the bistability jump is a genuinely new suggestion.

\section{MASS LOSS PREDICTIONS FOR THE MISSING LBVs}

In this section we present predictions of the mass-loss rates for
LBV-type stars with a luminosity typical for the LBV gap, i.e. with
$\logll = 5.7$. Previously, theoretical mass loss rates for LBVs have
been derived by Vink \& de Koter (2002) with the aim of investigating
the role of the bistability mechanism on the mass loss behavior of
these stars during their excursions across the HRD. Here we use the
same \mdot\ prediction method, which we briefly summarize below.

\subsection{Method}

In our method the mass-loss rate follows from tracking the radiative
energy loss of photons (to be precise: photon packages) traveling
through a model atmosphere in a Monte-Carlo simulation, and equating
the loss of photon energy to the gain in kinetic energy of the
outflowing gas. From an outside observer's point of view, this
decrease of radiative energy occurs when photons transfer momentum
(and energy) in interactions with moving ions in the flow.  An
iterative process is used to find the model in which the input mass
loss is equal to that found from the global radiative to kinetic
energy conversion. This \mdot\ value is the predicted mass-loss
rate. Details of the Monte-Carlo method are given in Abbott \& Lucy
(1985), de Koter et al. (1997), and Vink et al. (1999).

The model atmospheres used in the process are calculated using the
non-LTE code {\sc isa-wind}. The inner boundary of the atmosphere is
chosen to be at a Rosseland optical depth of $\sim$ 20 - 25, i.e.
sufficiently large to assure thermalization of the radiation
field. The density stratification gradually changes from being
dominated by the equation of hydrostatic equilibrium in the inner
regions, to being dominated by the stellar wind at and beyond the
sonic point (see below). The chemical species for which the statistical
equilibrium equations are solved explicitly are H, He, C, N, O, and
Si. The adopted abundances of these species account for surface helium
enrichment and CNO processed material, and are the same as used in
Vink \& de Koter (2002). For a full description of {\sc isa-wind}, we
refer the reader to de Koter et al. (1993, 1997).

Some important line driving elements, i.e. iron and to a lesser extent
sulfur (see also Sim 2004), are not explicitly accounted for but are
treated in a generalized form of the ``modified nebular
approximation'' described by Lucy (1987, 1999). This simplified
treatment may result in a (systematic) shift of the temperatures at
which the dominant ionization state of these species change relative
to full non-LTE calculations.  Indications that such an offset may
indeed exist have been presented by Vink et al. (1999). These authors
identify the occurrence of strong changes in the terminal velocity and
ionization of the winds of OB supergiants at spectral type B1
(corresponding to the bistability jump at $\sim$21\,000 K; Lamers et
al. 1995) to be a result of the transition of the dominant iron
ionization from Fe\,{\sc iv} to {\sc iii}. However, they predict this
jump to be at $\sim$ 25\,000 K, i.e. at an effective temperature that
is $\sim 4\,000$ K higher than the observed temperature of the
jump. Simulations over a wide range of wind densities show that the
Fe\,{\sc iv}/{\sc iii} ionization balance is mostly sensitive to
temperature, and not to density. Consequently, the predicted
bi-stability jump was found to comprise only the narrow range of $22.5
\leq \teff \leq 26.5$ kK (Vink et al. 2000). One may therefore expect
that for the somewhat denser winds of LBVs, the offset will be similar
to -- or slightly larger than -- the maximum B supergiant offset of
5\,500 K.  Indeed, for the LBV AG\,Car a temperature offset of about
6\,000 K was found (Vink \& de Koter 2002). We note that for O-type
stars, where Fe~{\sc iv} is dominant for all spectral sub-types, we
find good agreement between predicted and observed mass loss behavior
(see e.g. Vink et al. 2000; Benaglia et al. 2001; Herrero et al. 2002;
Repolust et al. 2004). As we anticipate that the modified nebular
approximation causes only an offset in the predicted temperature of
the bistability jump, we will apply a corrective shift $\Delta \teff =
-6\,000$\,K to our predictions.  This is intended only to guide the
reader in order to facilitate a meaningful comparison with the
observed bistability jump in B1 supergiants and in AG\,Car.

Our mass loss prediction method is distinct from the approach in which
the line force is parameterized in terms of force multipliers
(e.g. Castor, Abbott \& Klein 1975; Kudritzki et
al. 1989). The main advantages of our approach are the self-consistent
treatment of the ionization stratification and the fact that it
accounts naturally for multiple photon scattering processes. A
disadvantage of our method may be that we do not {\it de facto} solve
the momentum equation, i.e. the predicted mass loss values depend to
some extent on the adopted velocity stratification $v(r)$ above the
stellar photosphere. For the supersonic part of the wind we assume the
standard $\beta$-type velocity law, which below the sonic point
connects smoothly to the velocity structure implied by hydrostatic
equilibrium (see de Koter et al.\ 1997 for details).  A value of
$\beta=1$ has been found to yield excellent results in modeling the
spectra of O star winds (e.g. Puls et al.\ 1996), and is assumed here.
The predicted mass-loss rates are found to be rather insensitive to
this wind acceleration parameter for $\beta$ in the range 0.7 -- 1.5
(Vink et al.\ 2000). Note that a value of $\beta = 1$ has been
considered too low for dense winds; infrared observations by Barlow \&
Cohen (1977) indicate a more extended wind acceleration in
P~Cygni. Also, the extremely dense winds of Wolf-Rayet stars appear to
favor more gradual wind acceleration (i.e. larger value of $\beta$),
at least for the outer wind (see Hillier 2003 for a discussion).

We specified the terminal velocity by adopting values for the ratio
$\ratio $ of 1.3, 2.0, and 2.6.  Note that Lamers et al. (1995)
determined a ratio of 2.6 for Galactic supergiants of spectral type
earlier than B1, and a ratio of 1.3 for supergiants of spectral type
later than B1 (see also Kudritzki \& Puls 2000).

\subsection{Predictions of mass loss for LBVs with \logll=5.7}

We define a grid of models having input temperatures \tin\ between
11\,000 and 35\,000 K. For a fixed luminosity \logll=5.7, the input
radius then follows from the relation $L = 4 \pi R_{\rm in}^2 \sigma
\tin^4$. As the inner boundary is chosen to be deep in the
stellar photosphere (see above) this input temperature does not
reflect the actual effective temperature, nor does \rin\ reflect the
actual stellar radius. We define the stellar radius \rstar\ and
effective temperature \teff\ at the point where the thermalization
optical depth measured in the center of the photometric $V$ band (at
5555 \AA) equals $1/\sqrt{3}$ (see Schmutz et al.\ 1990 and de Koter
et al.\ 1996 for detailed discussions). Both \rstar\ and \teff\ are
therefore output quantities. For stars with relatively modest mass
fluxes, such as normal O stars, the winds will be optically thin and
\teff\ will only be marginally less than \tin. However, for LBVs,
which may lose mass at rates of $\mdot\sim 10^{-4}\,\msunyr$, there
may be a significant difference between these two temperatures. If the
wind is so strong that the visible light originates from layers near
or beyond the regime of rapid wind acceleration the star is considered
to have a ``pseudo-photosphere''. Wolf-Rayet stars show such optically
thick winds.

The formation of pseudo photospheres in LBVs may be favored by the
relatively low mass of these stars, as this will 1) increase the
photospheric scaleheight, and 2) lead to a larger mass loss (see
below). Stothers \& Chin (1996) estimate the mass of LBV stars with
\logll=5.7 to be roughly 17 \msun. As this value is uncertain we
consider masses that range between 10 and 25 \msun.

Predictions of LBV mass loss are presented in Figure~\ref{f_mdots} for
masses of 25, 20, 15, and 12 \msun.  Results are shown for three
different ratios of \ratio. The four panels clearly show that for
fixed temperature \tin\ the mass-loss rate increases with decreasing
stellar mass. Vink \& de Koter (2002) found $\log \mdot / \log M
\propto -1.8$ for masses 30 down to 10 \msun; these new results
agree. Note that for normal O stars the predicted slope is shallower
(i.e -1.3; Vink et al. 2000).  Also visible in Figure~\ref{f_mdots} is
a gradual shift of the location of the bistability jump towards higher
\teff\ for lower \ratio\ and stellar mass. This is because these
models have a higher wind density.  The Saha-Boltzmann ionization
equilibrium (as formulated in the modified nebular approximation) then
implies that recombination will occur at a somewhat higher
temperature.

Let us inspect the panel for $M$ = 15 \msun\ a bit closer, as this is
closest to the typical LBV mass derived by Stothers \& Chin for
\logll=5.7. Following the $\ratio=2.6$ curve the mass loss rate
increases by a factor of more than two at the predicted location of
the bistability jump between model~A at $\teff = 30$\,kK and model B
at 25\,kK. Given the observed characteristics of the bistability jump
(at $\sim$ 21 kK, see above), \ratio\ is expected to decrease from 2.6
to 1.3. Thus, switching over to the curve with \ratio\ = 1.3 (model~C)
the total increase in \mdot\ over the bistability jump is about a
factor 5. Even more relevant is that {\em the wind density increases
by a factor of $\sim$10}. These properties of the bistability jump
are similar to those predicted by Vink et al. (1999) and are explained
by changes in the line driving properties of iron (see \S\,4.1).

The key question is: {\em can this large increase in density make the
winds of LBVs optically thick and cause the formation of a
``pseudo-photosphere''?}

\subsection{The formation of pseudo-photospheres}

De Koter et al. (1996) assessed whether changes in wind properties
could explain the visual magnitude changes of $\Delta V \simeq 1$ to 2
mag observed in LBV stars.  On the basis of a parameter study they
concluded that pseudo-photospheres are unlikely to form in
LBVs. However, they did not investigate in detail the effect of an
order of magnitude change in the wind density of a star that is
extremely close to its classical Eddington limit at $\Gamma_e$=1,
where $\Gamma_e = \kappa L/4\pi cGM$ and $\kappa$ is due to pure
electron scattering. Figure~\ref{fig:bistab-mass} shows the result for
such calculations in terms of the change in effective temperature as a
function of stellar mass, for two \tin\ values. We verified, by
decreasing the wind density by an order of magnitude, that the
temperature \tin\ is fairly representative for the effective
temperature of the star {\em if the bistability jump would not
occur}. We see that the lower the mass, the closer the star gets to
its Eddington limit.

To gain insight into the wind optical thickness of the 15 \msun\ star
below the jump (A), and above the jump (C) (note that model B is an
intermediate step only), we plot the thermalization optical depth of
these models against their wind velocities in Figure~\ref{f_thz}.  The
hot model (A) only reaches the point where $\tau_{\rm thermz}$ exceeds
$1/\sqrt{3}$ in the photosphere, where the wind velocity is smaller
than 2 km/s. However, this situation is rather different for the
cooler model C with enhanced mass loss. In model C the thermalization
depth is reached at 19 km/s, i.e. at a velocity {\it above} that of
the sound speed (16 km/s).  This implies that model C starts to form
an optically thick wind, which may lead to the formation of a (modest)
pseudo photosphere.

Figure~4 explores the issue whether the size of the forming
pseudo-photosphere is large enough to create the LBV gap.  The top
panel for $\tin$ = 25 kK shows that for a current mass that is below
about 12 \msun\ (corresponding to $\Gamma_e = 0.8$) the star will
rapidly form an extended optically thick wind envelope as the star
crosses the bistability jump, reaching down to an effective
temperature of 13\,kK for a mass of 10.5\,\msun. For $\tin$ = 17.5 kK
a similar lower limit to the effective temperature is found, again for
$M = 10.5\,\msun$.  Though we tried to compute models for even lower
mass, these failed to converge as a result of their proximity to the
Eddington limit.

As stated above, an offset $\Delta \teff$ of about --6\,000 K should
be applied to have the predicted temperature of the bistability jump
match the observed value. Naively applying this shift would bring the
effective temperature at the low temperature side of the bistability
jump to about 7\,000 K, which agrees well with the location of the
yellow hypergiants. At about this temperature the extinction of a
gaseous medium reaches a maximum, so this may be expected to represent
the maximum redward shift such a star may achieve (Davidson 1987;
Appenzeller 1986). Therefore, for masses even lower than 10.5 \msun\
the stars would remain in the same region of the Hertzsprung-Russell
diagram.  Note that stars at the {\it observed} temperature of the
bistability jump have a core radius somewhat larger than do stars at
the {\it predicted} location of the jump. Therefore, the mass flux at
the high wind density side of the jump is expected to be only about
half of what we predict. This does not change our conclusions in a
significant way.

One may bring forward arguments that favor the formation of a pseudo
photosphere that is even more extended than predicted in our
calculations.  Both observational and theoretical indications exist
that at least some LBVs may show relatively slow wind acceleration,
i.e. a value $\beta > 1$ (Ignace et al. 2003). Though this does not
have a strong effect on the mass loss rate (see Vink et al. 2000), it
does lead to a further increase of the photospheric radius for a star
that already shows signs of pseudo photosphere formation (the
thermalization optical depth of such a model (model D; $\beta=1.5$) is
plotted in Fig.~\ref{f_thz}).  One could even envision the situation
that the initial formation of a (modest) pseudo photosphere may lead
to a runaway effect, i.e. a progressively slower velocity law
(equivalent to a progressively larger $\beta$). A test of this
scenario requires time-dependent hydrodynamic modeling of the outer
stellar envelope and wind, and is beyond the scope of this paper.

The above results suggest the following sequence of events: unstable
hot stars in the S\,Doradus instability strip below $\logll \simeq
5.8$ would find themselves positioned at the low temperature side of
the bistability jump. Compared to quiescent LBVs at higher luminosity,
their wind density would increase by up to an order of magnitude.
This may lead to the formation of a pseudo photosphere, pushing the
star toward cooler temperatures, and into the regime in which the
yellow hypergiants are observed, causing the gap in the S~Dor
instability strip presented in Figure 1.  Consequences of this
scenario are discussed in more detail below.

\section{DISCUSSION}

Proceeding with the assumption that the ``gap'' is a real feature of
LBVs on the HRD, we seek an explanation for why it may exist.  In this
paper we have hypothesized that the gap is a consequence of the
bistability jump --- namely, that at temperatures below about 21\,000
K, an abrupt change in the behavior of the line-driving mechanism
makes the winds of LBVs unstable enough that they may develop pseudo
photospheres and are then pushed toward cooler apparent temperatures
on the HRD.  For LBVs on the S Dor instability strip, the bistability
jump at 21\,000 K occurs at \logll$\simeq$5.8.

\subsection{Optically-Thick Winds and the Bistability Jump}

To test the above hypothesis, we have conducted numerical simulations
of line-driven stellar winds for LBVs at \logll=5.7.  Such a
luminosity is consistent with the evolutionary track of an initially
$\sim$40 \msun\ star (Maeder \& Meynet 1987, 1988; Stothers
\& Chin 1996). Our simulations show (Fig.\ 2) that the bistability
jump does play a significant role at \logll=5.7, for a range of
stellar masses, with higher mass-loss rates on the cool side of the
bistability jump. We find that for current masses above 20
\msun, LBVs do not form optically thick winds. In the mass range
between 15 and 20 \msun\ we find indications that they are on the
verge of forming pseudo photospheres or indeed that their atmospheres
become optically thick at the base of the wind. For even lower masses
LBVs form extended pseudo photospheres (Fig.\ 4).  Table 2 also gives
the classical electron-scattering Eddington ratio $\Gamma_e$ for each
model.  In LBVs, it has long been thought that an opacity-modified
Eddington limit may play a role, so that atmospheres become unstable
at 80\% or 90\% of the classical limit, instead of at $\Gamma_e$=1
(e.g. HD; Appenzeller 1989; Lamers \& Fitzpatrick 1988; Ulmer \&
Fitzpatrick 1998).  Indeed, for \logll=5.7 we find that severe pseudo
photospheres develop for $\Gamma_e \ga 0.8$.  This fact may give
critical clues to their evolutionary state, as discussed below in \S
5.2.  Additionally, our results show that on the cool side of the
bistability jump, stars with $\Gamma_e$ as low as $\sim$0.5 are on the
verge of having optically-thick winds.  Note that P\,Cygni, for which
the bistability mechanism was first introduced to explain the
variability of its stellar wind (which is not the same as forming a
true extended pseudo photosphere), has $\Gamma_e \simeq 0.5$.

For masses of 15-20 \msun, pseudo photospheres are quite weak, with
temperature shifts $\Delta T= T_{\rm in} - T_{\rm eff}$ of only about
1\,000 K at $M = 15\, \msun$. If LBV masses fall in this range for
\logll=5.7, then we would need to invoke a ``runaway'' effect to
explain the ``gap''.  As hinted earlier, the temperature drop of the
initial pseudo photosphere might cause a drop in the wind speed and
perhaps an increase in opacity, which in turn, might increase the
mass-loss rate, making the wind even more optically thick.  Similar
ideas have been discussed in the past with regard to the outbursts of
LBVs (Appenzeller 1986, 1987, 1989; Davidson 1987; HD). Perhaps a
different instability proposed in connection to the ``Yellow Void''
(de Jager 1998; see below) would then become relevant.  Presumably
such a runaway would continue until the temperature fell to about
7\,500 K (see Davidson 1987; Appenzeller 1986), when much of the
atmosphere would recombine and the opacity would drop. Again, this is
only relevant for relatively high LBV masses --- if a total of $\Delta
M \ga 25\,\msun$ can be shed by the star by the time it leaves the RSG
phase, then our calculations show that invoking such a ``runaway'' is
unnecessary.

The mass dependence (or rather, the L/M-dependence) of the instability
due to the bistability jump is probably the reason why other stars
seen in this region of the HRD are apparently immune.  For instance,
several B[e] stars reside near \logll=5.7 with spectral types of B0 to
B3 (Zickgraf et al.\ 1986).  Despite their high luminosity, these hot
supergiants show little photometric or spectroscopic variability,
indicating that they are not subject to the same instability that
affects LBVs.  Perhaps they have not yet lost enough mass to raise
their L/M ratios to a critical level.  It is interesting, however,
that the bistability jump may affect their complex, latitude-dependent
stellar winds (Zickgraf et al.\ 1986 Lamers \& Pauldrach 1991;
Pelupessy et al. 2000).

\subsection{The ``Missing'' LBVs as Post-RSGs}

Log$(L/L_{\odot})\simeq5.8$ is the luminosity where the S\,Doradus
instability strip intersects the bistability jump at $\sim$ 21\,000 K,
and it is also the limit above which no RSGs or YHGs are seen on the
cool side of the HRD (HD).  Is this just a coincidence, or perhaps,
are the two phenomena related?

We have shown that at $\logll \simeq 5.7$, LBVs on the S Dor
instability strip may develop pseudo photospheres because of the
heightened opacity in their atmospheres on the cool side of the
bistability jump --- but this only seems to be important for
sufficiently low masses such that $\Gamma_e >$ 0.5 (preferably
$\Gamma_e\ga$0.8), or for 10-15 \msun.  {\it This mass range agrees
well with that expected for a post-RSG of the same luminosity.}  For
an O star with initial mass 40 \msun\ that evolves directly to the LBV
phase, it is unlikely that its $\Gamma_e$ would be high enough to
allow the order of magnitude jump in wind density at the bistability
limit to form a significant pseudo photosphere. Instead, after heavy
mass loss during the RSG phase, a star with this initial mass would
have decreased to below 20 \msun\ (Maeder \& Meynet 1987, 1988) --- in
good agreement with our constraints.  This suggests strongly that the
``missing'' LBVs need to be post-RSG in order for our proposed
mechanism to work, and would thereby explain the apparent
``coincidence'' of the gap and the upper luminosity limit for RSGs.

{\it What about LBVs above and below the gap?}  While stars above
$\logll \simeq 5.8$ do not become RSGs, they are expected to lose
considerable mass, perhaps in an $\eta$ Car-like giant eruption
(Maeder 1989; Chiosi \& Maeder 1986; Stothers \& Chin 1999), before
becoming a classical LBV.  However, when they do finally settle down
to become normal quiescent LBVs on the S Dor instability strip, they
are on the hot side of the bistability jump where the wind density is
much lower.  The low-luminosity LBVs may be stable enough to exist on
the S Dor instability strip --- despite being on the cool side of the
bistability jump --- simply because their lower L/M ratios render the
bistability-induced pseudo photosphere mechanism inapplicable.  For
example, pseudo photospheres resulting from the bistability mechanism
only seem to be significant for $\Gamma_e > 0.5$, while the
low-luminosity LBVs typically have values of $0.2 \la \Gamma_e \la
0.5$.

\subsection{The Missing LBVs, the Yellow Hypergiants, and the ``Yellow Void''}

Yet another phenomenon that is important in this region of the HRD,
and may be related to the missing LBVs, is the so-called ``Yellow
Void'' (de Jager 1998; de Jager \& Nieuwenhuijzen 1997; Nieuwenhuijzen
\& de Jager 2000).  The Yellow Void is a region of instability in the
HRD between temperatures of roughly 7\,000 and 12\,500 K, and
luminosities of $5.4 \la \logll \la 6.0$.  Stars in this region can
have very low $g_{\rm eff}$ in their outer layers, and the sonic point
can be below photospheric levels.  In combination with stellar
pulsations this may lead to dynamically-unstable convective
atmospheres and very high mass-loss rates (de Jager; Stothers \& Chin
2001; Stothers 2003).  De Jager (1998) suggests that this instability
is only important for blueward-evolving post-RSGs (with high enough
L/M ratios), such as the cool hypergiants plotted in Figure 1.
Nieuwenhuijzen \& de Jager propose that several of the yellow
hypergiants (YHGs) are ``bouncing'' against the Yellow Void in an
attempt to evolve to warmer temperatures, but are stopped by the
instability and undergo high mass loss.

Are the YHGs somehow related to the missing LBVs?  A more direct and
provocative question might be to ask if the YHGs {\it are} the missing
LBVs --- in other words, if it were not for the pseudo photospheres
induced by the bistability jump or the Yellow Void, would the YHGs
reside on the S Dor instability strip and behave like LBVs?  The YHGs
occupy the same range of luminosities that seem to be missing from the
LBVs, and like the LBVs, they show conspicuous apparent temperature
variations at relatively constant bolometric luminosity
(Nieuwenhuijzen \& de Jager 2000; Figure 1).  They are the only stars
within this luminosity range that show such dramatic variability in
spectral type on short time scales (years to decades, like LBVs).

What is the relationship, if any, between this Yellow Void and the
bistability jump that we have addressed in this paper?  For masses
below 12 \msun, our proposed mechanism would have the same effect as
the Yellow Void --- i.e. it would halt blueward evolution at
$\sim$7\,500 K for post-RSGs.  For higher masses (15-20 \msun), we
have shown that the enhanced mass loss on the cool side of the
bistability jump may cause at most a modest pseudo photosphere to
develop at \logll$\simeq$5.7.  This alone might not be enough to push
a star far to the right on the HRD --- but it may be enough to push
the star into the Yellow Void, where a different instability then
takes over.  Likewise, the Yellow Void instability alone would not be
able to trigger this without the bistability jump, because the Yellow
Void only extends to temperatures as warm as $\sim$12\,500 K and does
not reach the S Dor instability strip.  Thus, for relatively high LBV
masses at \logll=5.7, these two mechanisms may work together to
account for the LBV gap.

An alternate view might be that the bistability jump is not so
important for explaining the missing LBVs, because they are unable to
get there anyway --- if they are indeed blueward-evolving post-RSGs, they
may simply be stopped by the Yellow Void and prevented from becoming
normal quiescent LBVs.  In that view, however, the bistability jump
may still play an important role in defining an extended blue boundary
of the Yellow Void.  This will be especially true if stars shed a
large amount of mass as a YHG, so that their L/M ratio is even higher
than immediately following the RSG phase.

IRC+10420 is particularly interesting in this regard.  It is at both
the upper luminosity boundary for cool stars, and is at the cool edge
of the Yellow Void (see Fig.\ 1).  It is a relatively unambiguous
case for a post-RSG, having OH masers despite its A-type spectrum (see
Oudmaijer 1998; Humphreys et al.\ 1997), and a shell that
qualitatively resembles the RSG nebula around VY CMa (Smith et al.\
2001).  Humphreys et al.\ (2002) suggest that IRC+10420 is at a
critical stage when it is just about to cross the Yellow Void and
emerge on the blue side as a hot LBV.  Although IRC+10420 has evolved
from a main-sequence star of roughly 40 \msun, its present-day mass is
most likely well below 20 \msun; Humphreys et al.\ (2002) favor 15
\msun, while Nieuwenhuijzen \& de Jager (2000) have suggested that its
current mass may even be as low as 6 \msun.  Thus, based on the
results of our calculations, if IRC+10420 were to evolve toward warmer
temperatures, we would expect it to be severely effected by the cool
side of the bistability jump, and it should form a pseudo photosphere.
In fact, our \logll=5.7 model at 10.5 \msun\ may be directly relevant
to IRC+10420. This model has a mass-loss rate of $2 \times 10^{-4}$
\msunyr, which is similar to the present-day rate of $\mdot
\simeq 1.5 \times 10^{-4} \msunyr$ for IRC+10420 deduced by
Humphreys et al.\ (2002), and $\mdot \simeq 5 \times 10^{-4} \msunyr$
measured fom CO lines by Oudmaijer et al.\ (1996). Humphreys et al.\
argue that such a rate is enough to cause a very dense and peculiar
wind with a pseudo photosphere and an apparent temperature of
$\sim$8\,500 K.

Of course these scenarios should be tested with quantitive
spectroscopy, but this is not an easy task in the regime of hydrogen
recombination, where the H$\alpha$ line is difficult to predict, and
because of close proximity to the Eddington limit. Quantitative
spectroscopy might result in a determination of $L$, \Teff, and \mdot\
for the yellow hypergiants, but arguably the most crucial paramater to
test our scenario is the stellar mass, which is unfortunately even
harder to obtain because of the degeneracy involved in simultaneously
determining both \mdot\ and the mass.  Because the yellow hypergiants
exhibit temperature variations, the approach outlined for AG~Car by
Vink \& de Koter (2002) could potentially be used. In this approach the
mass loss behavior as a function of time (and therefore temperature)
can be used to constrain the stellar mass.

\subsection{The evolution of massive stars at $5.6 \la \logll \la 5.8$}

The discussion above brings into question the still-uncertain
evolutionary sequences for massive stars with $\logll \la 5.8$.  These
stars are able to evolve to the right side of the HRD and go through a
RSG phase. Several investigators have discussed the idea that after
the RSG phase, these stars will then become low-luminosity LBVs and
eventually WR stars (i.e. O star $\rightarrow$ BSG $\rightarrow$ RSG
$\rightarrow$ LBV $\rightarrow$ WR; Maeder 1982; Chiosi \& Maeder
1986; Stothers \& Chin 1994, 1996).  For stars with $\logll \la 5.6$,
this seems reasonable based on Figure 1 and many other considerations.

However, for stars with \logll$\simeq$5.7$\pm$0.1, the situation may
be more complicated because of the interaction between the YHGs at the
Yellow Void, and their proximity to the upper luminosity limit for
cool stars.  Consider what will happen to IRC+10420 or $\rho$~Cas if
they are eventually able to cross the Yellow Void --- if they make it
to the blue side of the Yellow Void and become LBVs, they will find
themselves on the S Dor instability strip just to the cool side of the
bistability jump.  However, this is the location of the LBV ``gap''.
We have shown in this paper that in this zone of the HRD, post-RSGs
with $M \la 15 \msun$ (and especially post-YHGs with {\it even lower}
masses and higher values of $\Gamma_e$) will form pseudo photospheres
independent of the Yellow Void, and so it is likely that {\it these
stars will be pushed back across the Yellow Void!}  Hence, in order
for IRC+10420 or $\rho$~Cas to truly cross the Yellow Void, they must
make it all the way to the hot side of the bistability jump where the
wind density decreases by an order of magnitude.  Interestingly, this
is where we see several Ofpe/WN9 stars or candidate LBVs with
circumstellar ring nebulae, like S~119, Wra~751, and others.  These
Ofpe/WN9 stars have ring nebulae that expand at speeds of 20--30 km
s$^{-1}$, much slower than their stellar winds (Bianchi et al.\ 2004;
Pasquali et al.\ 1997, 1999; Hutsem\'{e}kers \& van Drom 1991).  In
fact, these nebular expansion speeds are closer to the wind speeds of
YHGs and RSGs, rather than the faster winds of O, WR, LBV, or even
Ofpe/WN9 stars themselves.

One reasonable interpretation, then, may be that the bistability jump
acts together with the Yellow Void, so that within a small range of
luminosity ($\logll \simeq 5.7 \pm 0.1$), post-RSGs will skip the LBV
phase, instead appearing only as YHGs before becoming slash stars and
eventually WR stars. (i.e. perhaps the YHG phase takes the place of
the LBV phase for these stars.)  Thus, from this line of reasoning one
might adopt the following evolutionary scenario for $\logll \simeq 5.7
\pm 0.1$:
\begin{displaymath}
  {\rm O star}      \rightarrow 
  {\rm BSG/B[e]}    \rightarrow
  {\rm RSG (OH/IR)} \rightarrow
  {\rm YHG}         \rightarrow
  {\rm Ofpe/WN9}    \rightarrow
  {\rm WR}.
\end{displaymath} 
The distinction between this scenario and a much more simplified one
(like O star $\rightarrow$ RSG $\rightarrow$ WR) is that these
comments apply to the apparent surface of the star, which at times is
a very thick pseudo photosphere.  During this entire post-RSG phase,
the core should continue to evolve blueward, independent of the star's
outer atmosphere.  This is why the star, after loosing essentially all
of its outer envelope, may eventually cross the Yellow Void (de Jager
\& Nieuwenhuijzen 1997) to appear on the hot side of the bistability
jump, perhaps as an Ofpe/WN9 star with a slowly-expanding ring nebula.

\acknowledgements \scriptsize

N.S.\ was supported by NASA through grant HF-01166.01A from the Space
Telescope Science Institute, which is operated by the Association of
Universities for Research in Astronomy, Inc., under NASA contract NAS
5-26555.  J.V.\ acknowledges financial support from 
the Particle Physics and Astronomy Research Council of the United Kingdom.

\begin{table}
\caption[]{Typical LBV models around the bistability jump
           for a present-day mass of 15 \msun.}
\label{t_models}
\begin{tabular}{ccccl}
\hline\hline
\noalign{\smallskip}
Model & \teff\  & log \mdot\ & \ratio\ & $\beta$  \\
\noalign{\smallskip}
      & (kK) & ($\msunyr$)  &          &          \\
\hline
\noalign{\smallskip}
A     & 30   &   -4.93      &  2.6     &   1      \\
B     & 25   &   -4.52      &  2.6     &   1      \\
C     & 25   &   -4.28      &  1.3     &   1      \\
D     & 25   &   -4.28      &  1.3     &   1.5    \\
\noalign{\smallskip}
\hline
\end{tabular}
\end{table}

\begin{table*}
\footnotesize
\caption[]{Predicted mass loss rates and effective temperatures.
           \teff\ corresponds to the total flux at the radius where
           thermalization optical depth in the photometric $V$ band
           at 5555 \AA\ equals $1/\sqrt{3}$.}
\label{t_apparant}
\begin{tabular}{lcccrcr}
\hline\hline
\noalign{\smallskip}
\noalign{\smallskip}
$M_*$   & log$L_*$ & $\Gamma_{\rm e}$  & $\ratio$ 
        & $\tin$ & log $\mdot$ & $\teff$ \\
\noalign{\smallskip}
 ($\msun$)  & ($\lsun$) &           &  & (kK) & $\msunyr$ & (kK) \\
\hline
25  & 5.7 & 0.383 & 2.6   &  35   & $-$5.23  & 34.8    \\  
    &     &  &       &  32.5 & $-$5.32  & 32.3    \\
    &     &  &       &  30   & $-$5.35  & 29.8    \\
    &	  &  &       &  27.5 & $-$5.37  & 27.3    \\
\noalign{\smallskip}
    &     &  & 1.3   &  25   &  $-$4.65 & 24.8    \\
    &	  &  &       &  22.5 &  $-$4.67 & 22.3    \\
    &     &  &       &  20   &  $-$4.67 & 19.8    \\
    &     &  &       &  17.5 &  $-$4.60 & 17.3    \\
    &     &  &       &  15   &  $-$4.55 & 14.8    \\
    &	  &  &       &  12.5 &  $-$4.60 & 12.3    \\
    &	  &  &       &  11   &  $-$4.71 & 10.8    \\
\noalign{\smallskip}
\hline
\noalign{\smallskip}

20  & 5.7 & 0.478 & 2.6   &  35   & $-$5.02  &  34.7   \\  
    &     &  &       &  32.5 & $-$5.14  &  32.2   \\
    &     &  &       &  30   & $-$5.18  &  29.7   \\
    &	  &  &       &  27.5 & $-$5.10  &  27.3   \\
\noalign{\smallskip}
    &     &  & 1.3   &  25   &  $-$4.48 &  24.7   \\
    &	  &  &       &  22.5 &  $-$4.51 &  22.2   \\
    &     &  &       &  20   &  $-$4.51 &  19.7   \\
    &     &  &       &  17.5 &  $-$4.41 &  17.2   \\
    &     &  &       &  15   &  $-$4.39 &  14.7   \\
    &	  &  &       &  12.5 &  $-$4.49 &  12.3   \\
    &	  &  &       &  11   &  $-$4.60 &  10.8   \\
\noalign{\smallskip}
\hline
\noalign{\smallskip}

15  & 5.7 & 0.638 & 2.6   &  35   & $-$4.76  &  34.3   \\  
    &     &  &       &  32.5 & $-$4.88  &  31.9   \\
    &     &  &       &  30   & $-$4.93  &  29.4   \\
    &	  &  &       &  27.5 & $-$4.70  &  27.0   \\
\noalign{\smallskip}
    &     & 0.559 & 1.3   &  25   &  $-$4.28 &  23.9   \\
    &	  &  &       &  22.5 &  $-$4.30 &  21.7   \\
    &     &  &       &  20   &  $-$4.25 &  19.3   \\
    &     &  &       &  17.5 &  $-$4.21 &  16.8   \\
    &     &  &       &  15   &  $-$4.26 &  14.4   \\
    &	  &  &       &  12.5 &  $-$4.38 &  12.0   \\
    &	  &  &       &  11   &  $-$4.48 &  10.6   \\
\noalign{\smallskip}
\hline
\noalign{\smallskip}

12  & 5.7 & 0.797  & 2.6   &  35   & $-$4.53  & 33.3    \\  
    &     &  &       &  32.5 & $-$4.62  & 31.1    \\
    &     &  &       &  30   & $-$4.62  & 28.6    \\
    &	  &  &       &  27.5 & $-$4.27  & 26.4    \\
\noalign{\smallskip}
    &     & 0.698  & 1.3   &  25   &  $-$3.98 & 19.9    \\
    &	  &  &       &  22.5 &  $-$4.00 & 18.5    \\
    &     &  &       &  20   &  $-$4.02 & 17.0    \\
    &     &  &       &  17.5 &  $-$4.08 & 15.6    \\
    &     &  &       &  15   &  $-$4.18 & 13.8    \\
    &	  &  &       &  12.5 &  $-$4.32 & 11.6    \\
    &	  &  &       &  11   &  $-$4.42 & 10.2    \\
\noalign{\smallskip}
\hline
\noalign{\smallskip}

11.5 & 5.7 & 0.729 & 1.3   &  25   &  $-$3.89 & 17.9 \\

\noalign{\smallskip}
\hline
\noalign{\smallskip}

11  & 5.7 & 0.762 &  1.3   &  25   &  $-$3.79 & 15.5 \\

\noalign{\smallskip}
\hline
\noalign{\smallskip}

10.5 & 5.7 & 0.798 & 1.3   &  25   &  $-$3.67 & 12.9  \\ 
\noalign{\smallskip}

\hline
\end{tabular}
\normalsize
\end{table*}

\begin{figure*}
\epsscale{0.95}
\plotone{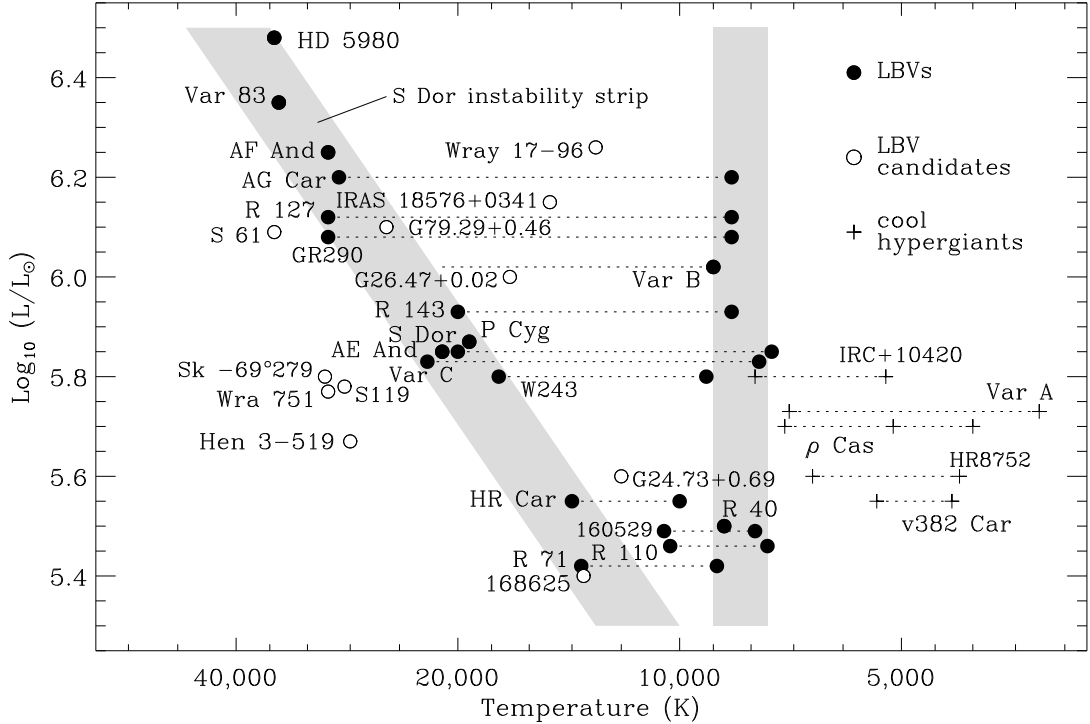}
\caption{The upper H-R Diagram for LBVs and related stars, including
LBV candidates with circumstellar shells and cool hypergiants.  Shaded
areas represent the places where LBVs are most commonly found: either
on the diagonal S Dor instability strip, or the nearly
constant-temperature outburst phase.  Values for L and T are adopted
from HD or de Jager (1998), except as noted here for indivudual stars:
HD~5980 (Koenigsberger et al.\ 1998); Var 83, AF And, and AE And
(Szeifert et al.\ 1996); Wray 17-96 (Egan et al.\ 2002); AG Car (HD;
Hoekzema et al.\ 1992; Lamers et al.\ 1996a); IRAS 18576+0341 (Clark
et al.\ 2003b); G79.29+0.46 (Higgs et al.\ 1994); S 61 (Pasquali et
al.\ 1997); Romano's Star GR290 (Polcaro et al.\ 2003); P Cyg (Lamers
\& De Groot 1992; Lamers et al.\ 1996b; HD); S Dor (Massey 2000; HD);
W243 (Clark \& Negueruela 2004); Sk -69$\arcdeg$279 (Thompson et al.\
1982); S 119 (Crowther \& Smith 1997); Wra 751 (Hu et al.\ 1990); Hen
3-519 (Smith et al.\ 1994); Var A (Humphreys et al.\ 1987);
G24.73+0.69 and G26.47+0.02 (Clark et al.\ 2003a); and HD 168625 (van
Genderen et al.\ 1992).}
\end{figure*}

\begin{figure*}
\centerline{\psfig{file=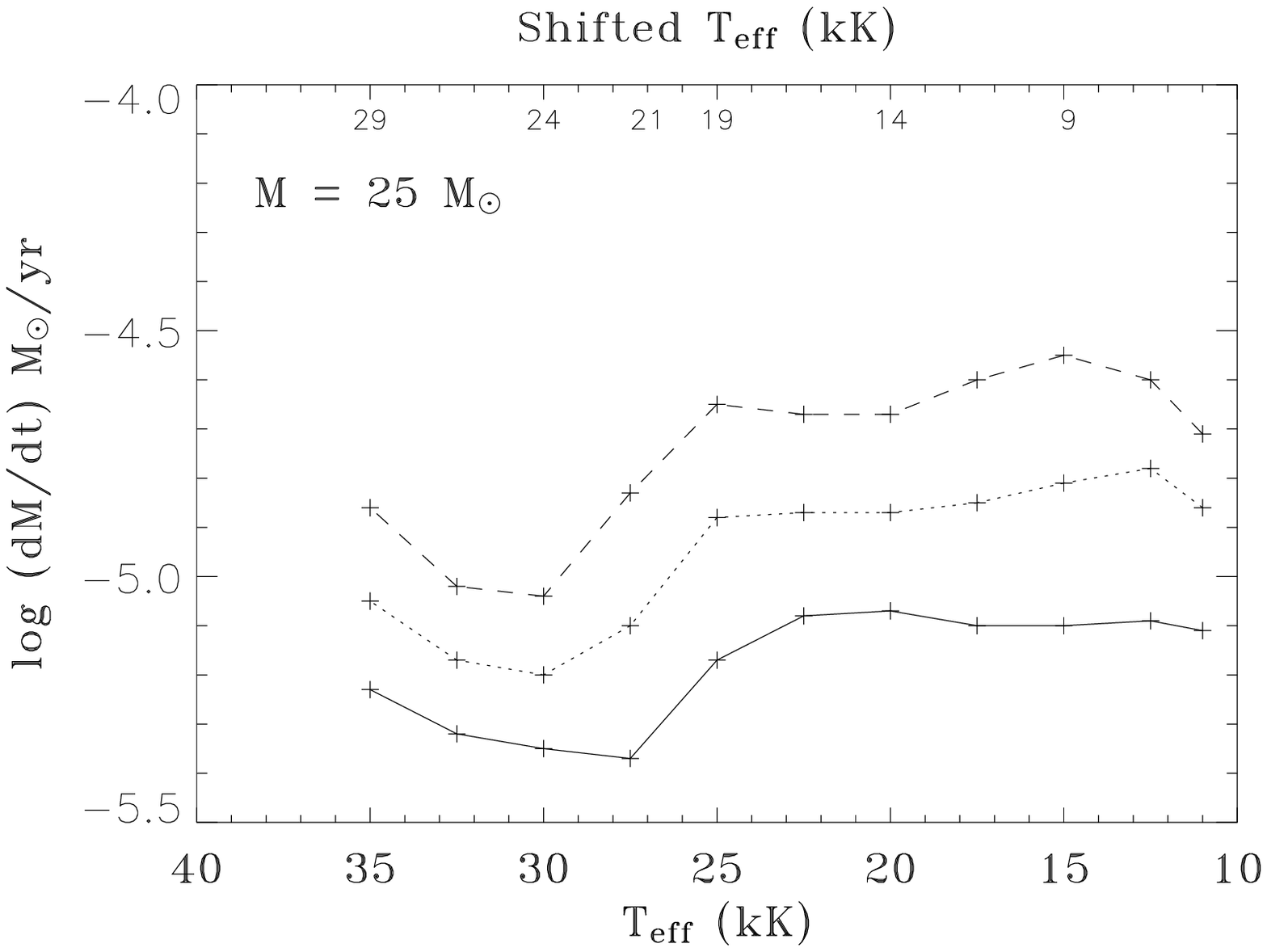, width = 8 cm}}
\centerline{\psfig{file=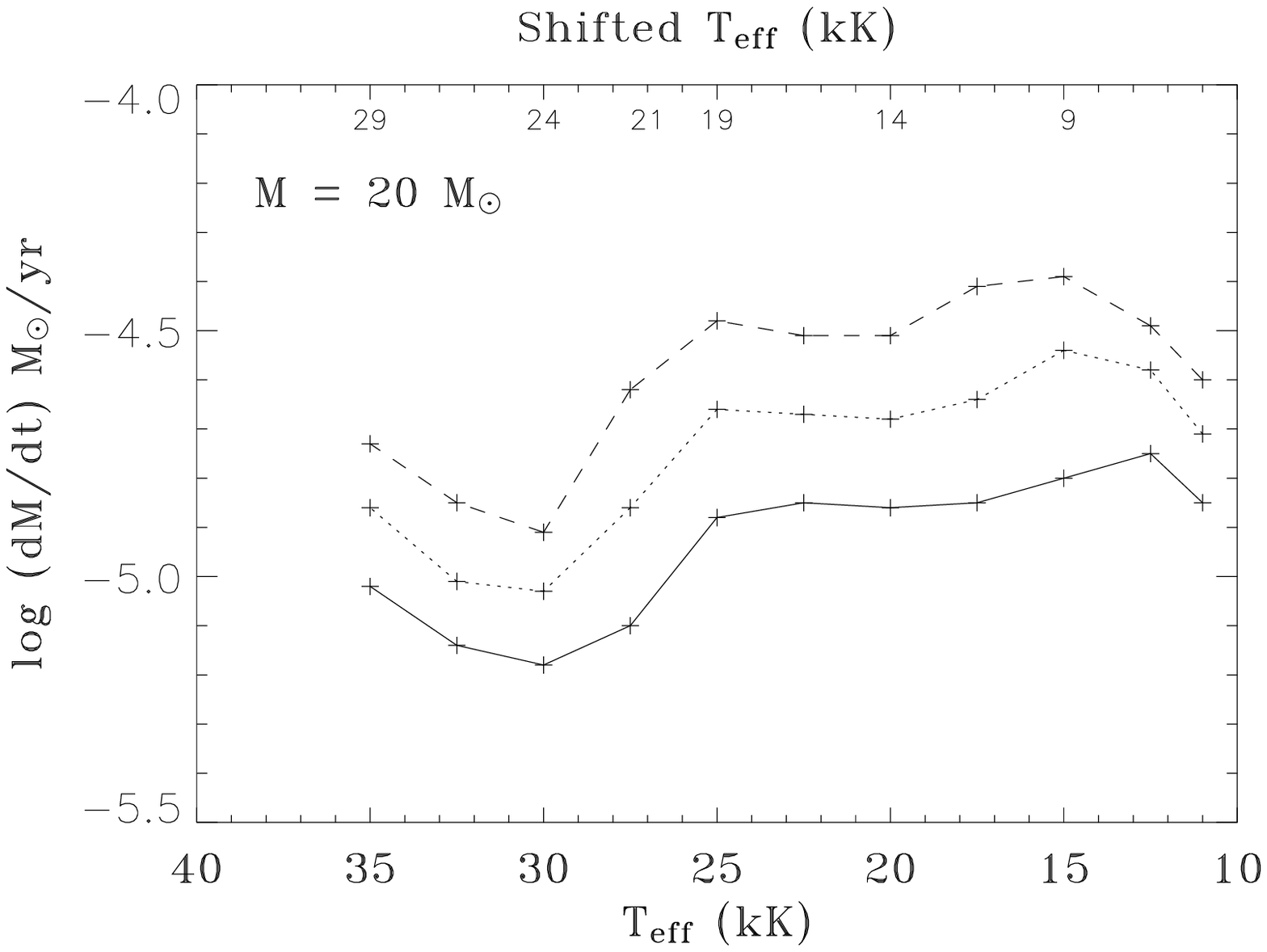, width = 8 cm}}
\centerline{\psfig{file=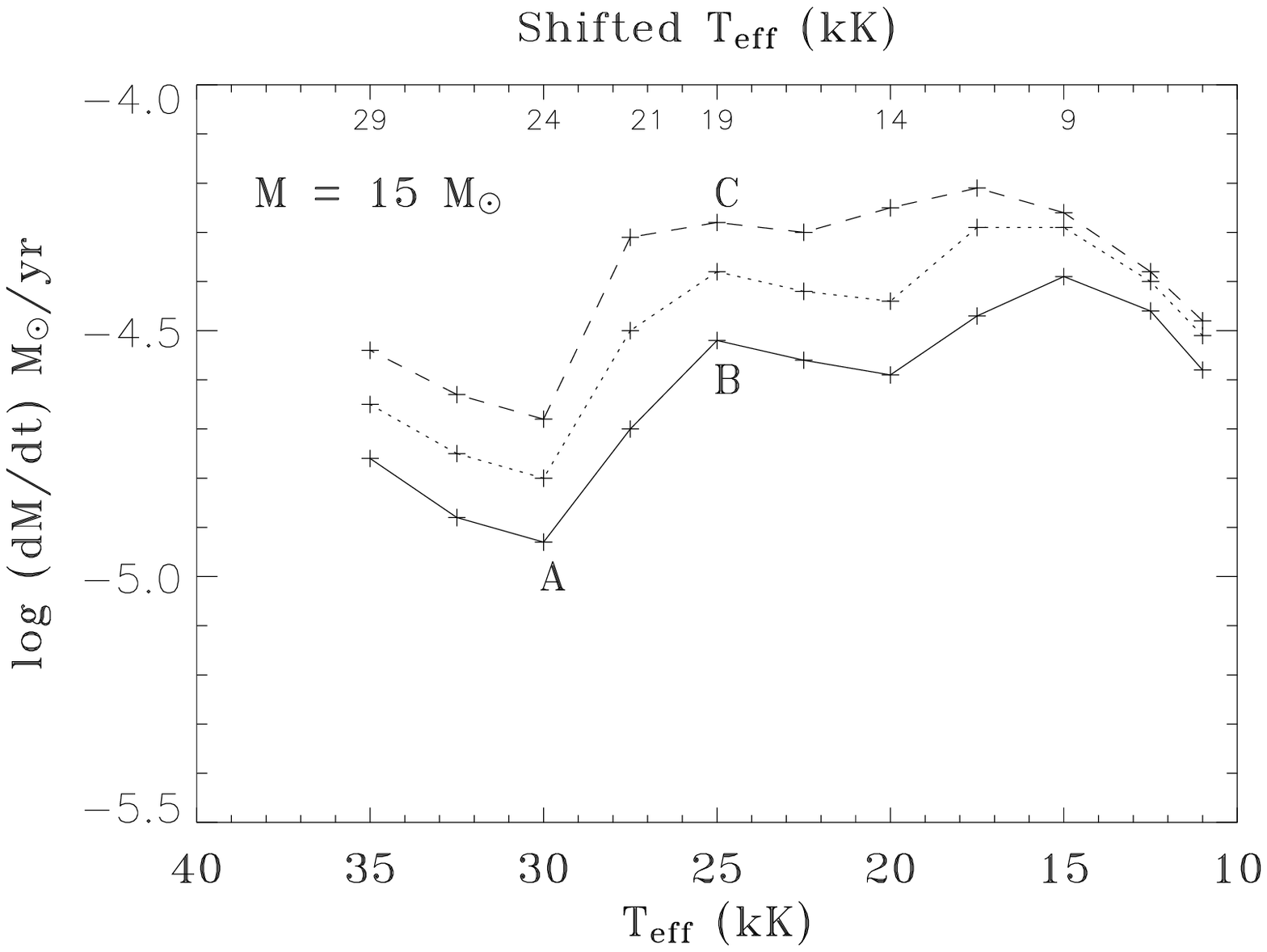, width = 8 cm}}
\centerline{\psfig{file=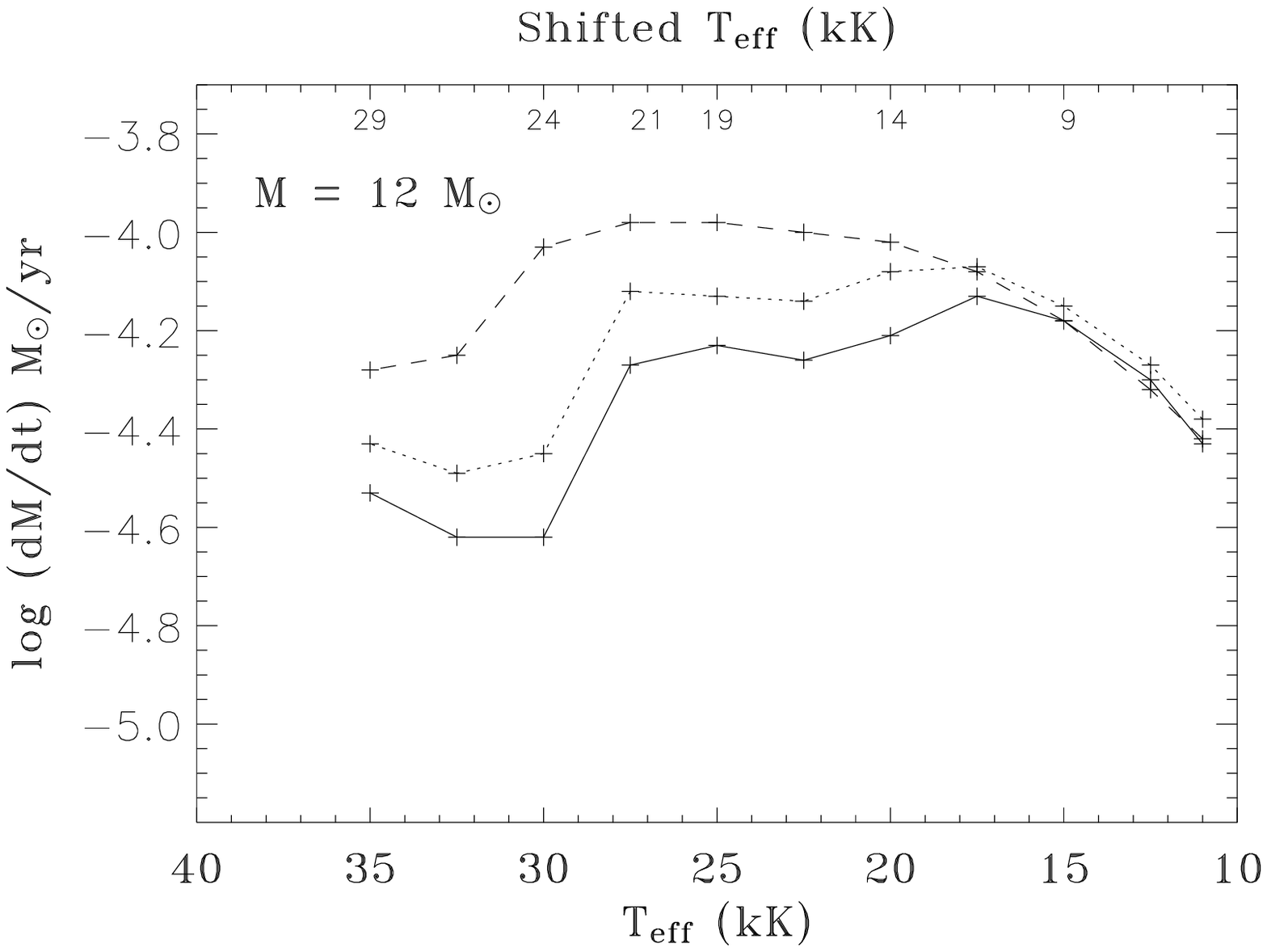, width = 8 cm}}
\caption{Predicted mass-loss rates as a function of effective temperature for
         four adopted stellar masses indicated in the upper left
         corners.  The solid lines represent the models with \ratio\ =
         2.6, the dotted lines are for the intermediate \ratio\ = 2.0,
         and the dashed lines are for the \ratio\ = 1.3 models
         representing the cool side of the bistability jump.  All
         models have \logll\ = 5.7.  Each panel also shows the shifted
         value for \Teff\ after applying the --6000 K correction (see
         text); this is only intended to guide the reader in
         interpreting effects due to the location of the bistability
         jump.}
\label{f_mdots}
\end{figure*}

\begin{figure*}
\centerline{\psfig{file=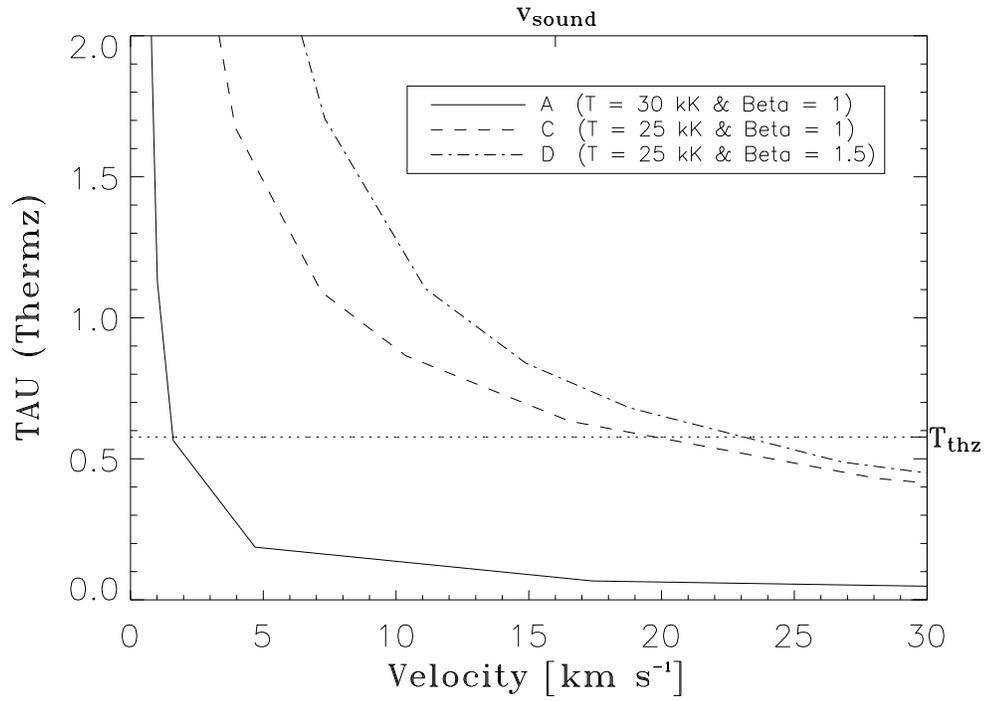, width = 14 cm}}
\caption{The thermalization optical depth against the wind velocity for 
         three typical models (A, C and D) for an LBV of 15 \msun\
         (see Figure 2$c$).  Models C and D on the cool side of the
         bistability jump appear to have modest psuedo photospheres.}
\label{f_thz}
\end{figure*}


\begin{figure}
\centerline{\psfig{file=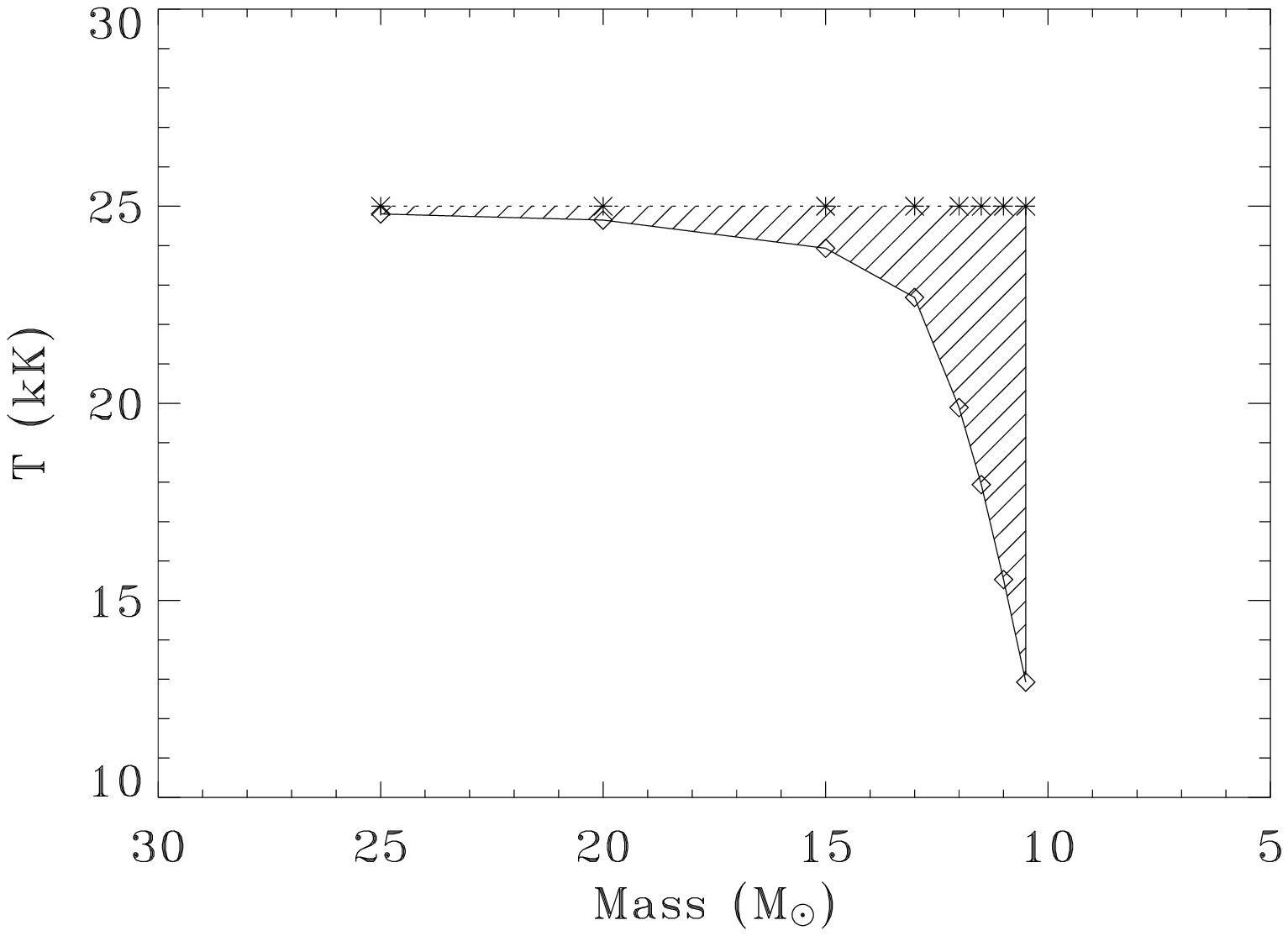, width = 12 cm}}
\centerline{\psfig{file=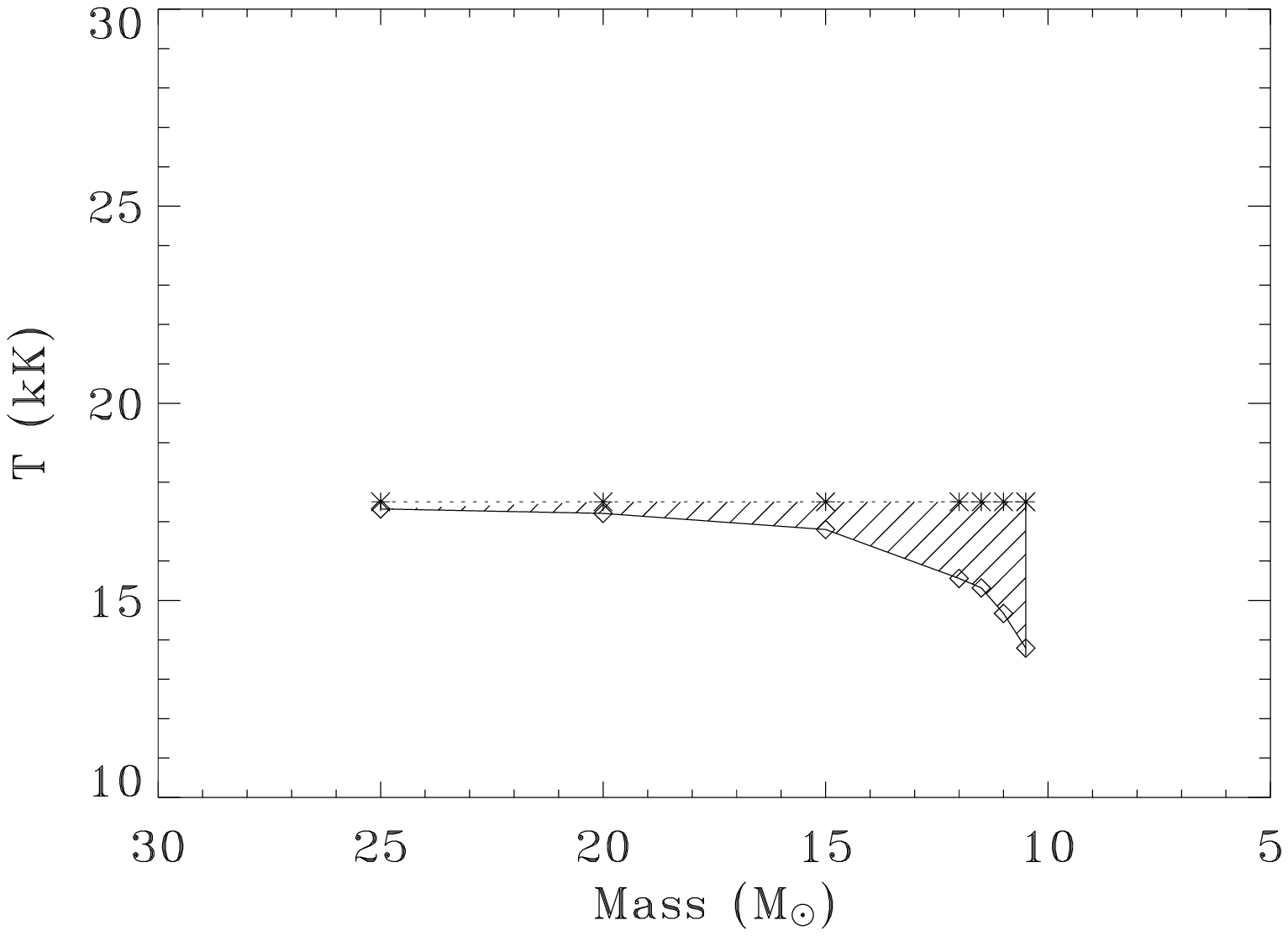, width = 12 cm}}
   \caption{The change in effective temperature as a result of
            crossing the bi-stability limit -- causing an order of
            magnitude change in wind density -- as a function of
            stellar mass for a star with $\logll = 5.7$. The top
            diagram shows models for $\tin = 25\,000$\,K (denoted by
            $\star$ symbols); the bottom diagram for $\tin =
            17\,500$\,K.  The calculated effective temperature of the
            star is denoted by a $\diamond$.  When the stellar mass
            becomes as low as $M \sim 10.5\,\msun$ the star forms a
            significant pseudo photosphere, resulting in an effective
            temperature of about 13\,000\,K for both \tin\
            cases. Applying the corrective shift of $6\,000$\,K (see
            text) this becomes $\teff \sim 7\,000$\,K, which
            corresponds to the location of the yellow hypergiants.
            This indicates that blueward evolving supergiants that
            have lost significant mass during a prior red supergiant
            phase may be susceptible to pseudo photosphere formation.
            } \label{fig:bistab-mass}
\end{figure}

\end{document}